# AN LLM -POWERED ASSESSMENT RETRIEVAL-AUGMENTED GENERATION (RAG) FOR HIGHER EDUCATION

Reza Vatankhah Barenji [1*], Nazila Salimi [2], Sina Khoshgoftar [1]

[1] Department of Engineering, School of Science and Technology, Nottingham Trent University, Nottingham, UK

[2] School of Construction, property and surveying, College of Technology and Environment, London South Bank University, London, UK

reza.vatankhahbarenji@ntu.ac.uk

## Abstract

Providing timely, consistent, and high-quality feedback in large-scale higher education courses remains a persistent challenge, often constrained by instructor workload and resource limitations. This study presents an LLM-powered, agentic assessment system built on a Retrieval-Augmented Generation (RAG) architecture to address these challenges. The system integrates a large language model with a structured retrieval mechanism that accesses rubric criteria, exemplar essays, and instructor feedback to generate contextually grounded grades and formative comments. A mixed-methods evaluation was conducted using 701 student essays, combining quantitative analyses of inter-rater reliability, scoring alignment, and consistency with instructor assessments, alongside qualitative evaluation of feedback quality, pedagogical relevance, and student support. Results demonstrate that the RAG system can produce reliable, rubric-aligned feedback at scale, achieving 94–99% agreement with human evaluators, while also enhancing students' opportunities for self-regulated learning and engagement with assessment criteria. The discussion highlights both pedagogical limitations, including potential constraints on originality and feedback dialogue, and the transformative potential of RAG systems to augment instructors' capabilities, streamline assessment workflows, and support scalable, adaptive learning environments. This research contributes empirical evidence for the application of agentic AI in higher education, offering a scalable and pedagogically informed model for enhancing feedback accessibility, consistency, and quality.

Keywords: Education, Large Language Model (LLM), Assessment, Retrieval-Augmented Generation (RAG), GenAI

## 1. Introduction

Feedback has traditionally been conceptualised as information that allows learners to compare their actual performance with desired standards or goals [1]. It is widely regarded as one of the most influential factors in shaping student learning and achievement [2]. Over the past two decades, the role of feedback has become increasingly prominent, particularly as higher education has expanded its emphasis on formative assessment practices designed to improve instruction and academic outcomes [3]. Despite this recognition, higher education continues to face persistent concerns regarding the adequacy, quality, and timeliness of feedback provision, especially in large-scale courses where instructors struggle to provide personalised and meaningful responses [3]. Consequently, student satisfaction with assessment and feedback remains consistently lower than with other aspects of their studies [4].

For feedback to contribute effectively to learning, students must be able to clearly understand learning goals, compare their current performance with those goals, and take informed action to improve [3-4]. A shift toward learner-centred pedagogies does not diminish the teacher's role; rather, it repositions educators as designers of effective learning environments and facilitators of meaningful interactions that support students' engagement with feedback processes [4]. However, within contemporary higher education settings characterised by large enrolments, educators face substantial challenges in providing individualised assessment and detailed feedback for all students [5]. These pressures are exacerbated when traditional assessment formats such as standardised tests are prioritised for administrative efficiency rather than deep learning [6]. As a result, alternative strategies such as peer feedback [6] and self-assessment [4] have been introduced to distribute evaluative responsibilities and support learning. Yet these methods require considerable scaffolding, and when teaching teams are constrained by time or resources, students often experience reduced support, limiting the impact of feedback on their academic development [7].

Importantly, a transition away from teacher-centred practices does not automatically guarantee effective student engagement with feedback. Many learners continue to struggle to recognise their active role in interpreting, applying, and seeking feedback [8]. Essay writing has long been valued as a method for assessing higher-order cognitive skills, critical reasoning, and individual expression [9]; however, the scoring of essays is frequently criticised for its subjectivity and variability among human markers. This has driven research into automated and data-driven scoring methods, including neural networks and coherence models [10], statistical classifiers such as Naïve Bayes and Random Forest [11], and natural language processing techniques such as semantic and sentiment analysis [12]. Large-scale educational organisations have responded similarly: EdX uses Automated Essay Scoring (AES) technologies, whereas Coursera relies on structured peer-review systems [13]. Although these approaches each offer strengths, they also have limitations and are most effective when combined. Still, concerns remain that feedback may simply "dangle" before learners without fostering genuine change [14, p. 121], underscoring the importance of re-evaluating how the responsibilities for giving and receiving feedback are shared between students and educators.

The introduction of generative AI (GenAI) into educational practice has intensified these debates. Following the release of ChatGPT in November 2022, discussions have emerged regarding the reliability of AI-generated content, its implications for assessment integrity and student learning, and the broader impact on educators' professional roles. As a result, education is currently a highly contested domain in relation to the adoption of GenAI [15-16].

Nevertheless, educators continue to apply professional judgement and creativity to develop instructional materials, and GenAI have been increasingly explored as tools to support this work [17]. Research demonstrates that GenAI can assist in generating learning resources, assessment questions, and instructional content [18-19]. When prompted effectively, such models can rapidly produce large volumes of educational material [19]. However, due to their probabilistic nature, their outputs must be treated as provisional suggestions rather than authoritative content [20]. GenAI have also been tested for evaluating or scoring student responses, yet their role remains supportive rather than substitutive [21]. Ethical, methodological, and pedagogical considerations require educators to integrate these tools cautiously[18]. Despite these limitations, GenAI can strengthen feedback cycles by enabling iterative, timely interactions that help students meet learning outcomes [17-18].

The effectiveness of such systems, however, depends critically on the reliability of their outputs. While GenAI can generate assessment materials at scale, they often default to tasks focused on recall rather than higher-order thinking [22]. Without genuine semantic understanding, their potential for supporting evaluation and feedback of complex cognitive skills remains limited [22]. Consequently, researchers recommend restricting GenAI use to formative or low-stakes assessment contexts where high-quality training data are available [18-19].

In response to these challenges, Retrieval-Augmented Generation (RAG) has emerged as a promising framework that enhances GenAI performance through the integration of external, curated knowledge sources [23]. Unlike standalone GenAI, which rely solely on pre-trained parameters, RAG combines a generative model with a retrieval system that dynamically accesses relevant documents, enabling responses grounded in up-to-date and context-specific information [23-24]. RAG can interpret contextual cues, retrieve targeted materials, and generate adaptive outputs that reflect domain-specific reasoning[24]. This hybrid architecture addresses common limitations of GenAI notably hallucination, outdated knowledge, and inconsistency while improving transparency and reliability [23-24]. In educational environments, RAG systems can retrieve lecture notes, rubrics, exemplars, and academic resources from institutional repositories, providing personalised explanations, tailored feedback, and adaptive support for learners. By merging generative adaptability with structured retrieval, RAG offers a scalable and pedagogically grounded strategy for deploying GenAI in assessment, tutoring, and personalised learning.

Although RAG technologies have advanced rapidly, their adoption within educational research and practice remains limited [25-27]. Existing studies overwhelmingly focus on conventional GenAI tools such as ChatGPT. To the best of the authors' knowledge, applications of RAG in teaching, learning, and assessment are under-explored, despite its potential to overcome key limitations of GenAI -only systems. By grounding model outputs in curated, domain-specific knowledge, RAG significantly enhances accuracy, consistency, and alignment with assessment criteria.

The present study addresses persistent challenges in providing timely, high-quality, and scalable feedback in higher education by developing and evaluating a GenAI-powered agentic assessment system grounded in a Retrieval-Augmented Generation (RAG) framework. Integrating a large language model with structured prompt engineering and a curated corpus of rubric criteria, exemplar essays, and lecturer feedback, the system autonomously retrieves relevant assessment knowledge and generates contextually aligned scores and formative comments. The system was evaluated using a mixed-methods approach, combining quantitative analyses of scoring reliability, inter-rater agreement, and alignment with human-generated grades with qualitative assessment of feedback quality and pedagogical relevance. Findings from its deployment across 701 student essays demonstrate that the RAG system can deliver consistent and rubric-faithful feedback at scale, significantly reducing turnaround time while maintaining close alignment with human judgement. At the same time, the study highlights important pedagogical considerations including the need for transparency, human oversight, and support for students' feedback literacy underscoring that AI systems function most effectively as partners rather than replacements for educators. By demonstrating both the practical feasibility and pedagogical implications of the developed assessment RAG, this research contributes to the growing field of AI-enhanced learning technologies and offers evidence-based insights into how such systems can improve feedback accessibility, support self-regulated learning, and strengthen teaching capacity in increasingly resource-constrained educational environments.

## 2. Literature and background

From a traditional transmission perspective, feedback has often been viewed as a linear process in which information and advice flow from an expert to a novice [3, 28]. This approach, however, tends to place limited responsibility on students for making feedback effective [1,4]. Higher education policies have frequently emphasised what teachers should do when providing feedback, while neglecting the role of students in interpreting and using it effectively. Consequently, feedback that is generic, delayed, or non-actionable has been criticised for limiting students' engagement and learning [29]. Scholars argue that both teachers and students must share responsibility in feedback processes, as effective feedback depends on active dialogue and mutual understanding [30]. When feedback is conceptualised as part of an interactive process, it becomes a powerful trigger for improving performance [31].

Effective feedback is multifaceted, and its success depends on a range of attributes, including timeliness, responsiveness, and relevance to learners' needs [2,4]. There is no universal feedback model suitable for all learners or contexts [32], and its impact often depends on how well it bridges the gap between current and desired performance [2]. Feedback can be summative, focusing on evaluation and outcomes, or formative, focusing on reflection and improvement [32]. Formative feedback encourages learners to take ownership of their learning by engaging in self-regulated processes that drive continuous improvement [32]. Taras (2009) clarified that formative assessment combines summative evaluation with feedback that learners can understand and apply, making it feasible even in large-scale educational settings [33]. Effective feedback not only informs learners of their progress but also challenges assumptions and supports self-regulated learning [2-4]. Recent studies highlight the growing importance of self- and peer-assessment, which allow students to evaluate their own performance critically and build confidence and metacognitive awareness [34-35].

With the expansion of large-scale and online education, providing individualised, high-quality feedback has become increasingly challenging [36]. AI has therefore emerged as a promising tool to support assessment and feedback processes. Although AI integration in education offers significant potential, scholars caution against overestimating its capabilities and emphasise the need for sustainable, human-centred implementation [37]. In particular, Automated Essay Scoring (AES) systems have demonstrated how AI can emulate aspects of human evaluation by analysing linguistic, syntactic, and semantic features in student writing [38]. AES systems, such as E-rater and IntelliMetric, employ natural language processing (NLP) and deep learning to provide reliable scores and constructive feedback. They can detect writing patterns, identify weaknesses, and offer suggestions for improvement, thereby enhancing learning outcomes [38]. AI-based assessment tools have been shown to provide adaptive, individualised feedback with minimal human intervention, reducing educators' workload while maintaining pedagogical quality [39]. Despite the promise of commercial AES systems, they are often closed-source, role-based, and limited in their adaptability for pedagogical personalisation [39].

Early studies on GenAI and assessment following the release of ChatGPT examined ability to perform academic tasks such as answering multiple-choice and open-ended examination questions [40]. Others explored whether AI-generated content could be reliably distinguished by humans or detection tools. While findings vary, there is general consensus that GenAI demonstrates adequate competence in passing professional examinations and producing coherent academic texts. These developments highlight both opportunities and challenges for higher education assessment [41]. On the one hand, GenAI enables new possibilities such as automated feedback generation, essay scoring, and personalised assessment design [42]. On the other hand, it raises significant concerns about academic integrity, given students can submit AI-generated work undetected and current detection tools remain unreliable [43]. Further issues include inequitable access to GenAI tools, data privacy and bias in AI algorithms, limited AI literacy among staff and students, and the spread of misinformation. Moreover, the absence of clear guidelines on GenAI-assisted assessment creates ambiguity around what constitutes academic misconduct [42]. For instance, should students using GenAI for language refinement be penalised, and should non-AI-assisted work be viewed as more authentic? Such questions underscore the need to revisit core principles of assessment particularly validity, fairness, and security and to redesign assessment practices that prepare students for learning and evaluation in an AI-mediated academic environment [42].

Despite their growing use in educational assessment, GenAI present several shortcomings in both qualitative and quantitative dimensions of essay evaluation [18-20]. From a quantitative perspective, GenAI -based scoring approaches often struggle with reliability and consistency, particularly when evaluating diverse writing styles, linguistic variations, or culturally nuanced arguments[18]. The absence of transparent scoring algorithms further complicates validity checks and reproducibility, leading to concerns about fairness and bias across student cohorts [20]. Qualitatively, GenAI lack genuine interpretive understanding, critical judgment, and awareness of the discipline-specific context delivered within a course attributes that human assessors naturally apply when evaluating originality, argumentation, and reflective depth. Their performance also depends heavily on the quality of the data provided and the prompts used during operation [42]. Since students' essays often follow different structures and may not always align with specified assessment criteria, ensuring consistency and validity across large cohorts remains a significant challenge.

While GenAI-based systems can serve as powerful tools for adaptive assessment, their educational effectiveness ultimately relies on deliberate human oversight, well-designed instructional frameworks, and robust quality-assurance processes to maintain fairness and pedagogical integrity at scale [18]. Although GenAI-generated feedback may be linguistically coherent, it often overlooks subtle rhetorical strategies, conceptual nuances, or the learner's developmental trajectory [40,42]. Furthermore, GenAI tend to emphasize surface-level features such as grammar and structure over higher-order reasoning, producing feedback that appears polished but pedagogically shallow [20]. These limitations underscore the need for human or human-like oversight, rigorous model calibration, and hybrid evaluation frameworks that combine automated precision with human interpretive insight in essay assessment. Developing open, explainable, and transparent GenAI -based agents capable of evaluating student work in a human-like manner therefore represents a highly valuable advancement for education.

Retrieval-Augmented Generation (RAG) represents an innovative paradigm within agentic artificial intelligence, combining the generative capabilities of GenAI with dynamic retrieval of external knowledge to enhance response accuracy and relevance [24]. Unlike conventional GenAI that rely solely on pre-trained parameters, RAG systems function as autonomous agents capable of accessing curated knowledge bases such as rubrics, exemplar essays, or instructional resources and integrating this information into the generation process. This dual mechanism of retrieval and generation allows RAG to produce outputs that are both contextually grounded and semantically coherent, addressing common limitations of standalone GenAI such as hallucinations, outdated information, and lack of domain specificity [24].

In educational contexts, RAG-based systems have the potential to address several persistent challenges in assessment and feedback [25-27]. First, they can enhance scalability by autonomously processing large numbers of student submissions, providing timely and consistent evaluations that would be infeasible for instructors alone. Second, by leveraging curated knowledge and rubrics, RAG can support validity and alignment, ensuring that feedback is linked to clearly defined learning outcomes and assessment criteria. Third, the agentic nature of RAG allows for adaptive, personalized feedback, guiding students based on their specific writing style, conceptual understanding, and prior performance, which promotes self-regulated learning and reflective practice [26-27].

Moreover, RAG addresses qualitative limitations of traditional GenAI feedback by incorporating structured, domain-specific knowledge, reducing reliance on superficial features such as grammar and style, and providing more pedagogically meaningful guidance [26]. It also mitigates consistency issues in scoring, particularly in large classes, by standardizing feedback outputs while maintaining the flexibility to adapt responses based on retrieved information. By combining automated precision with human-like interpretive reasoning, RAG can complement instructor efforts, ensuring that both formative and summative assessments are informative, reliable, and actionable [27].

## 3. Methodology

### 3.1. Research design

This study adopted a mixed methods research design organized as a closed-loop sequence. The primary objective was to develop, implement, and evaluate an RAG contains GenAI -powered agent and Retrieval System to automate the retrieve knowledge to generation of feedback and grades for student submissions. In the first phase, the retrieval system for the RAG was constructed and all assessment-related materials provided to the students and course team, including learning objectives, assessment plan, grading rubric, assignment feedback template are stored in the developed retrieval system by the power of GenAI. These materials contained detailed marking criteria and learning outcomes serving as foundational knowledge for the agent. In the second phase, the GenAI-powered agent was developed to assess student submissions, producing both formative feedback and summative grades. To enhance control and reduce model variability, prompt engineering techniques were employed to restrict the GenAI's outputs strictly to the information contained within the provided knowledge base. This ensured that all feedback and grades produced by the agent were contextually relevant, criterion-referenced, and consistent with the intended learning outcomes. In the third phase, a mixed methods approach was employed to evaluate the system's performance. Quantitative analyses were used to compare RAG generated scores with instructor-assigned grades to determine the level of agreement, consistency, and accuracy. Qualitative data, derived from instructor reflections on the RAG-generated feedback, were analysed to explore perceptions of usefulness, clarity, and fairness. This triangulation provided a holistic understanding of the system's capacity to enhance learning and support feedback processes.

### 3.2 Participants and Data

This study was conducted within the undergraduate engineering faculty of a university, focusing on a first-year core course titled *Introduction to Engineering*. A total of 732 engineering students from eight different programmes were enrolled in the course, which was delivered in an online learning format due to the large class size and limited instructor availability. The course aims to introduce students to the fundamentals of engineering practice, professional communication, and reflective problem-solving through applied written exercises. As part of the assessment, each student was required to complete four essays over the semester, resulting in 2,730 essay submissions in the academic year as 701, 683, 680, and 666 submissions in each of the lots.

The course engages with engineering-related themes such as engineering design and innovation, sustainability, ethics, and teamwork. The written assignments were designed not only to assess technical understanding but also to promote reflective and analytical thinking, which are essential competencies in engineering education. Each essay ranged between 800 and 1,000 words and was aligned with course weekly topics. Assessment plans were made available to students through the course's virtual learning environment, outlining submission timelines and grading criteria. To support student learning, sample essays from previous cohorts categorised as strong, moderate, and weak examples were shared to illustrate performance expectations. Within the cohort, 29 students were formally registered as having specific learning needs or neurodiverse profiles, and appropriate learning accommodations were provided in accordance with institutional policy.

To facilitate the assessment and feedback process, the course team developed detailed assessment rubrics (Table 1) outlining clear criteria for evaluating each assessment component, including its associated sub-elements. These rubrics were also shared with students as part of the assessment plan to promote transparency and support student understanding of grading expectations. Formative feedback from the course team was required to be provided within three weeks of assessment submission. To ensure consistency in evaluation, two instructors independently scored and provided feedback on a representative subset of student essays using the rubric. This process supported efficient yet reliable evaluation across a large volume of submissions. Inter-rater agreement was calculated using Cohen's Kappa [44] to assess the degree of scoring consistency between lecturers. Essays with conflicting or inconclusive ratings were jointly reviewed and re-evaluated to reach consensus. The resulting lecturer-graded subset of essays was then uploaded into the retrieval component of the GenAI-powered RAG system as part of the knowledge base. This served as reference material for evaluating the performance of the agent developed in this study. The system was assessed on its ability to provide nuanced, constructive, and human-like feedback across many submissions, thereby demonstrating its scalability, reliability, and pedagogical alignment within a high-enrolment engineering education context.

Table 1. Assessment rubric for one of the assessment elements

| Criteria | Excellent (80–100%) | Good (65–79%) | Satisfactory (50–64%) | Needs Improvement (0–49%) |
|---|---|---|---|---|
| 1. Problem Definition and Understanding (20%) | Demonstrates clear and comprehensive understanding of the engineering problem; articulates key parameters, | Clearly defines the problem and objectives; identifies main parameters but with minor | Defines the problem with limited detail; some parameters or constraints are | Problem poorly defined or misunderstood; lacks clarity or relevance to |
| | constraints, and objectives with strong contextual awareness. | omissions in context or scope. | missing or unclear. | engineering context. |
| 2. Application of Engineering Principles (25%) | Applies relevant mechanical engineering concepts and principles accurately and effectively; integrates theoretical knowledge with practical reasoning. | Demonstrates understanding of relevant principles with minor errors or gaps; links theory to practice adequately. | Shows basic understanding of principles but limited application or incorrect interpretation in some areas. | Misapplies or omits fundamental engineering principles; lacks evidence of conceptual understanding. |
| 3. Design Approach and Methodology (25%) | Presents a logical, innovative, and well-structured design process; justifies all design choices using sound reasoning and appropriate tools or models. | Uses a clear and structured design approach with appropriate justification for most choices; some reasoning may lack depth. | Describes a design process but lacks clear structure or justification for key decisions. | Design approach unclear, poorly structured, or unjustified; lacks evidence of systematic problem solving. |
| 4. Critical Reflection and Evaluation (15%) | Provides insightful evaluation of design outcomes; identifies limitations and proposes realistic improvements or alternative approaches. | Reflects critically on outcomes; discusses limitations with some relevant suggestions for improvement. | Offers basic evaluation of results; reflection is descriptive rather than analytical. | Lacks reflection or evaluation; fails to recognize limitations or learning outcomes. |
| 5. Communication and Presentation (15%) | Report is exceptionally well-organized, concise, and professional; uses technical language accurately; includes well-labeled figures, tables, and references. | Report is clear and coherent; technical language mostly accurate; minor issues in organization or formatting. | Report communicates ideas adequately but lacks clarity or contains formatting and language issues. | Report poorly structured, unclear, or unprofessional; major errors in technical communication or referencing. |

### 3.3 Proposed GenAI-Powered RAG system

The GenAI-powered RAG system was developed using n8n (https://n8n.io/ ), a workflow automation platform that enables seamless integration of multiple data sources, Application Programming Interfaces (APIs), and computational processes in a modular and scalable manner. The RAG framework was designed to process and evaluate large volumes of student essay submissions while generating adaptive, rubric-aligned feedback. The developed system consists of two main subsystems. The first subsystem focuses on building a robust retrieval component, which ingests knowledge from course-team–provided documents and makes this information available to the agent when requested. The second subsystem involves the development of the assessment agent, which uses the retrieved knowledge to evaluate essays and produce feedback consistent with human grading practices. Both subsystems are powered by GenAI, enabling human-like processing, reasoning, and decision-making throughout the assessment workflow.

The workflow of the retrieval system was designed to efficiently access, index, and deliver relevant content from the knowledge repository whenever a new essay submission was processed. The system utilizes a vector-based retrieval mechanism implemented with Supabase, while Google Drive serves as the platform for uploading new files. Upon receiving a PDF essay, the GenAI-powered agent generates contextually grounded responses based on the retrieved materials. The retrieval component ensures that the RAG is provided with the most relevant knowledge, and the generative component synthesizes this information into actionable scores and feedback comments.

As shown in Figure 1, the retrieval system workflow begins with the ingestion of input files in PDF format, stored in a centralized Google Drive folder. These PDFs may include course-team–generated resources such as rubric guidelines, exemplar essays from previous cohorts, or coded feedback used for training purposes. The workflow is configured to automatically detect newly uploaded PDFs in the Google Drive folder. Once detected, the content of each PDF is extracted using optical character recognition and converted into text. Metadata such as file name, upload date, and document type are also captured to support efficient indexing and retrieval. The extracted text is then passed to a vectorization module that transforms the textual content into high-dimensional numerical representations (embeddings). These embeddings encode semantic meaning, enabling the system to measure similarity between new student submissions and existing documents. All embeddings along with their associated metadata and original text are stored in a Supabase vector database. This vector-based repository supports semantic search, allowing the retrieval system to return the most contextually relevant documents when queried by a student essay. By using vector search rather than keyword-based retrieval, the system can identify relevant rubric criteria, exemplar essays, or instructor feedback even when the wording differs between the stored resources and the student's submission. When a new essay is submitted, the agent queries the vector database for documents with the highest semantic similarity. The top-k results (e.g., the five most relevant documents) are then passed to the generative component to produce contextually informed feedback.

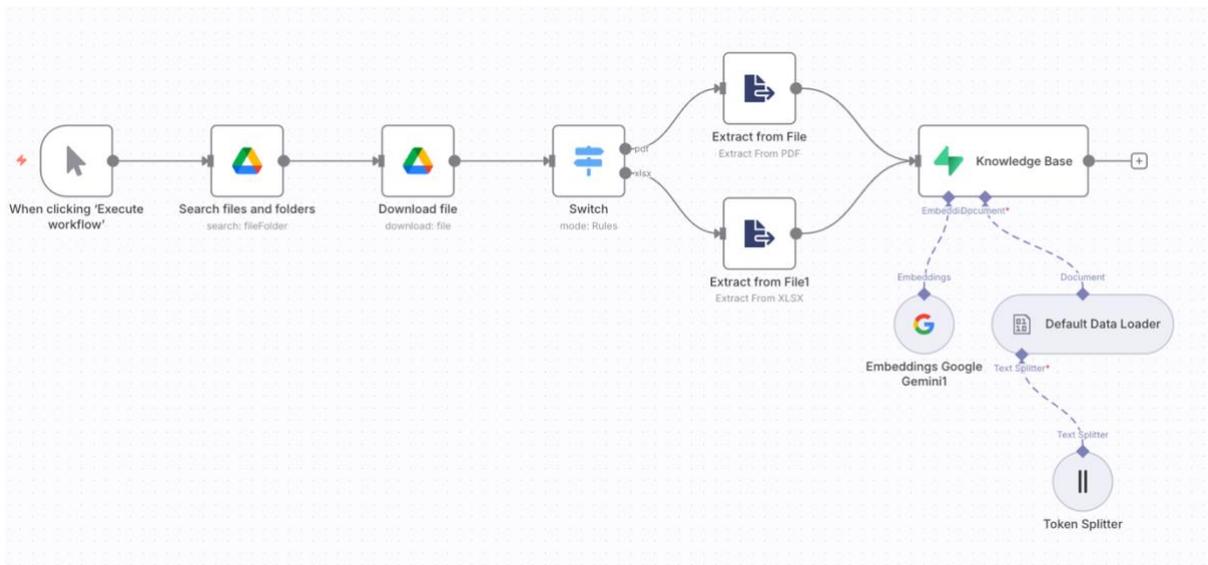

Figure 1 Workflow of the developed retrieval system

The workflow of the assessment agent is illustrated in Figure 2. Student essays, submitted as PDFs, are uploaded to a centralized Google Drive folder. The agent initiates the assessment process via a dedicated input request from the student within the workflow. Upon submission, the essay text triggers a query to the Supabase vector repository, and the retrieval component returns the most relevant rubric criteria, exemplar essays, and instructor comments. This ensures that the GenAI has a context-rich set of references before generating feedback.

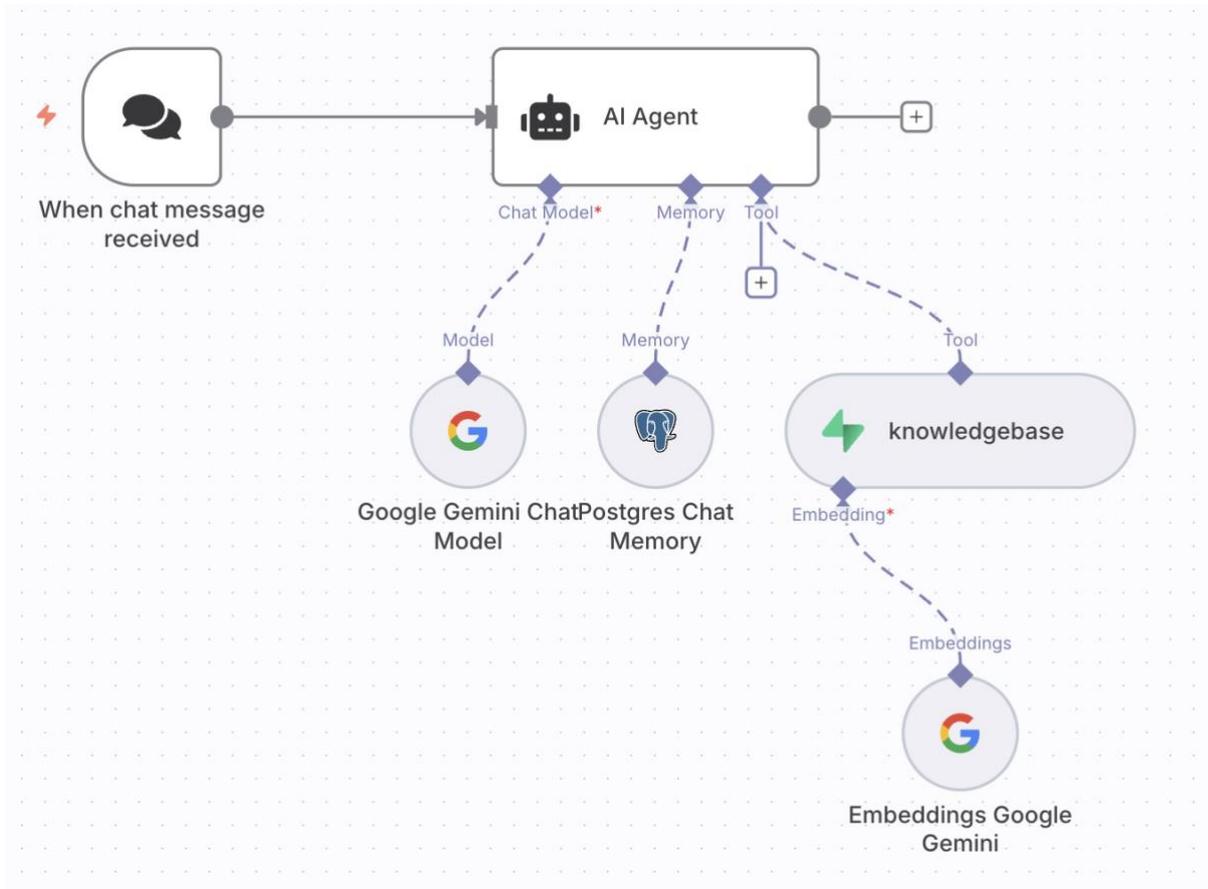

Figure 2 Workflow of the developed Assessment Agent

The retrieved documents are combined with a structured prompt information containing the essay text, relevant rubric guidelines, and examples of instructor feedback. Prompt engineering is applied to ensure that the GenAI's output remains aligned with assessment objectives and maintains consistency across multiple essays. Using a GenAI model (Gemini in this case) integrated into the workflow via an API node, the agent generates both a numerical score and qualitative feedback. The feedback synthesizes insights from the retrieved documents, adapting previous instructor examples to the specific content and quality of the new submission.

The generated scores and feedback are formatted and stored in a database and can optionally be sent to course team for review. This allows human-in-the-loop validation, ensuring that any anomalies or inconsistencies are corrected before feedback is released to students. Course team approved feedback can then be re-ingested into the vector knowledge repository, expanding the system's knowledge base and improving retrieval accuracy for future essays. This iterative process enables the RAG system to learn over time while maintaining alignment with pedagogical standards.

## 4. Results

A total of 701 student essays were analysed within a large-scale undergraduate engineering course. This dataset provided a robust context for evaluating the performance, reliability, and pedagogical validity of the developed RAG-based assessment system. The system's functionality was examined both qualitatively (e.g., depth of feedback, alignment with rubric descriptors) and quantitatively (e.g., scoring reliability, inter-rater agreement, variance explained). The RAG architecture integrated a vector-based retrieval subsystem, which located rubric elements, exemplar essays, and instructor-generated feedback, with an assessment agent responsible for generating automated scores and qualitative comments. Together, these components enabled pedagogically aligned and contextually informed evaluation across a large volume of submissions.

4.1 Phase 1: Human Baseline Scoring and Reliability Analysis

In the first phase, 50 essays were independently assessed by two lecturers to establish a baseline measure of human scoring consistency. Each essay was graded using the course rubric (see Table 1 and Table 2) and accompanied by qualitative comments addressing content accuracy, structure, critical reflection, and reasoning.

Inter-rater reliability among the human assessors was evaluated using Cohen's Kappa, yielding $\kappa = 0.63$, $p < .001$, which corresponds to a moderate–substantial level of agreement according to Landis and Koch (1977) [45]. To further assess scoring alignment, an Intraclass Correlation Coefficient (ICC) was calculated using a two-way random-effects model, resulting in $ICC(2,1) = 0.71$, 95% CI [0.58, 0.82], $p < .001$, indicating good reliability and consistency across the raters.

This result demonstrates good reliability, suggesting that differences between human raters were systematic and could be normed through calibration. Where discrepancies occurred, consensus meetings were held and scoring disagreements were resolved, producing a refined, high-quality human-coded dataset. These agreed-upon ratings were ingested into the RAG system as ground-truth reference material.

Table 2 Assessment Rubric for One Assessment Element

| Criterion | Performance | Approx. % Range | Weight | Weighted Contribution |
|---|---|---|---|---|
| 1. Problem Definition | Excellent | 90% | 20% | 18.0 |
| 2. Engineering Principles | Satisfactory | 60% | 25% | 15.0 |
| 3. Design Approach | Good | 75% | 25% | 18.75 |
| 4. Reflection & Evaluation | Needs Improvement | 45% | 15% | 6.75 |
| 5. Communication | Excellent | 90% | 15% | 13.5 |
| Total Weighted Score | | | | 72.0% |

4.2 Phase 2: Expansion of Human Reference Corpus

During Phase 2, the instructors independently graded an additional 100 essays (50 per instructor). This expanded corpus included a broader range of writing quality to improve representation and reduce bias in the knowledge base. The combined dataset of N = 150 instructor -scored essays (50 from Phase 1 and 100 from Phase 2) was indexed in the retrieval system. Each entry included: full essay text, numerical score, qualitative feedback, and associated rubric descriptors. This dataset served as the training and retrieval backbone for the RAG system.

Descriptive statistics for the human-scored corpus indicated a mean score of 71.4% (SD = 9.62), with scores approximately normally distributed (Shapiro–Wilk W = 0.97, p = .09) and ranging from 42% to 92%. These results confirmed that the dataset provides a representative and reliable referential distribution for calibrating and validating automated scoring within the RAG system.

4.3 Phase 3: Initial RAG-Based Scoring and Human Validation

The RAG system was then used to score an additional 100 essays. For each submission, the retrieval subsystem identified the most relevant rubric descriptors, exemplar essays, and prior lecturer feedback. The assessment agent synthesised this retrieved knowledge to generate; a numerical score, item-level qualitative comments, and a holistic feedback summary. Two instructors independently reviewed the RAG-generated outputs.

4.3.1 Quantitative Alignment Between Human and RAG Scores

A comparison between instructor scoring and RAG-generated scores revealed strong alignment between human and AI evaluations. Pearson correlation analysis indicated a high degree of association (r = 0.89, p < .001), while the mean absolute error (MAE) was 2.94 percentage points and the root mean square error (RMSE) was 3.62 percentage points, demonstrating minimal deviation from human scoring. Bland–Altman analysis showed a mean difference of –1.8%, indicating that the RAG system was slightly stricter than lecturers, with 95% limits of agreement ranging from –8.2% to +4.6%. Collectively, these results suggest that the RAG system produces scores that are highly consistent with human assessors, with error margins well within acceptable pedagogical limits.

### 4.3.2 Instructor Approval of RAG Feedback

Instructors evaluated the appropriateness of RAG-generated feedback, approving 94% of cases for inclusion in the knowledge base. The few disagreements that arose were primarily related to the tone of feedback either being perceived as too generic or overly directive and to instances where specific rubric criteria were insufficiently elaborated. Table 3 provides an illustrative example comparing lecturer feedback with the RAG-generated outputs, highlighting the system's alignment with human evaluative standards.

Table 3. Example of Quantitative Grading by Lecturers and RAG

| Criterion | Instructor Feedback | RAG-Based AI Feedback | Comparison / Observation |
|---|---|---|---|
| 1. Understanding of Ethical Issue | You identified the ethical problem clearly and linked it to sustainability and professional responsibility. Consider referencing formal codes of ethics for depth. | Demonstrates solid understanding of the ethical issue, highlighting professional responsibility. Explicit reference to formal codes of ethics is suggested. | Both highlight understanding; instructor feedback is more encouraging, while RAG emphasizes formal reference—complementary insights. |
| 2. Application of Ethical Theories | Good application of utilitarian and deontological perspectives. Could explore how each theory leads to different decisions. | Ethical theories are applied accurately; lacks detailed comparison of utilitarian vs. duty-based reasoning. | Instructor focuses on reasoning depth; RAG focuses on theoretical comparison. Both point to deeper analysis needed. |
| 3. Critical Reflection | Reflected on tension between commercial pressures and ethical responsibility. Consider including how personal values shape decisions. | Awareness of ethical tension is present but reflection is descriptive; suggest adding personal/professional implications. | Instructor feedback is more personalized; RAG feedback is concise but identifies same reflection gap. |
| 4. Communication and Structure | Well-organized essay; some sentences could be simplified for clarity. | Clear and coherent writing; minor wordiness, concise conclusion recommended. | Both note clarity issues; RAG is slightly more formulaic. |
| 5. Presentation & Language | Professional language with minor grammar errors. Proofread for consistent tense. | Language is professional; minor grammatical inconsistencies noted. | Both note small errors; RAG is more rigid in suggestions. |
| Overall Comment | Strong essay demonstrating understanding of ethics; deepen integration of theory with real-world examples for distinction. | Solid demonstration of ethical understanding; recommend deeper theoretical integration and context-specific reflection. | Instructor provides motivational, developmental tone; RAG is accurate but generic. |
| Grade | 78% | 75% | RAG aligns closely with instructor grade; |

| | | | slightly stricter on theory depth. |
|---|---|---|---|

4.4 Phase 4: Second Evaluation Cycle and System Refinement

In a subsequent round, 100 additional essays were assessed by the improved RAG system. Following human review, 99% of the RAG-generated outputs were approved, indicating a high level of accuracy and reliability. Reviewer feedback highlighted enhanced contextualisation of the feedback, closer alignment with the instructor's writing style, and minimal need for manual correction. These improvements reflect the cumulative effect of iterative human-in-the-loop refinement of the retrieval corpus, demonstrating that the system can progressively adapt and enhance its pedagogical fidelity over successive assessment cycles.

4.5 Summary of System Performance

Across all evaluation phases, the RAG assessment system demonstrated high scoring accuracy, with a correlation of $r = 0.89$ and a mean absolute error of less than 3%, and showed strong alignment with human feedback patterns. The system's reliability increased following iterative refinement, and it proved scalable to large cohorts, including 701 student essays, while consistently adhering to rubric descriptors. The final implementation indicated that the RAG system can effectively emulate lecturer-style evaluation at scale without compromising pedagogical fidelity. Importantly, the context-grounded feedback produced by the system supports self-regulated learning by enabling students to interpret their scores, identify performance gaps, and monitor their improvement trajectories throughout the semester.

5. Discussion

The findings of this study illustrate both the promise and the pedagogical challenges associated with deploying a RAG system for large-scale essay assessment. While the system demonstrated strong reliability particularly its high level of scoring consistency with human assessors and generated feedback that aligned well with rubric criteria, the results also highlight several limitations that affect its pedagogical applicability and long-term integration into teaching practice.

A key limitation concerns the extent to which pedagogical judgement can be embedded within an automated, retrieval-driven model. Although the system retrieved high-quality rubric descriptors, exemplars, and lecturer-generated feedback, human markers routinely draw on tacit disciplinary knowledge and interpretive judgement that extend beyond codified criteria. The results indicate that while the RAG system can emulate structural features of effective feedback, it occasionally missed deeper conceptual nuances or holistic interpretations that an experienced instructor would identify. This suggests that automated feedback though consistent may be less responsive to complexities such as subtle argumentation quality, conceptual originality, or disciplinary reasoning.

The findings also raise pedagogical concerns regarding potential narrowing of student learning. Because the RAG system generates feedback closely aligned to existing exemplars and rubric phrasing, there is a risk that students may begin writing to "fit the model," optimising their essays to satisfy predictable algorithmic expectations. This could lead to formulaic writing and reduced intellectual risk-taking an issue particularly salient given the system's strong consistency, which may make its feedback appear authoritative. Such patterns are pedagogically problematic, as disciplinary learning requires students to experiment, synthesise diverse perspectives, and develop their own academic voice.

Another important limitation, evidenced in student feedback logs and system output reviews, is the one-directional nature of the RAG-generated responses. Although the RAG system produced clear and structured comments, it could not engage in dialogic clarification or personalised guidance both essential components of high-quality feedback. Students receiving automated comments did not have an immediate mechanism to ask questions or negotiate interpretations, potentially constraining their ability to meaningfully act on feedback. This highlights a pedagogical gap between receiving feedback and understanding how to apply it effectively.

Moreover, the study identified varying levels of student readiness to interpret RAG-generated feedback. Some students applied the comments effectively, while others demonstrated misunderstandings or difficulty prioritising suggestions. This variation suggests that automated feedback may require scaffolding such as in-class modelling, peer review activities, or follow-up discussions to ensure equitable uptake. Without such supports, AI-driven feedback may inadvertently exacerbate differences in student confidence, academic literacy, and self-regulation skills.

Bias within the knowledge base also remains a concern. Although efforts were made to curate diverse high-quality exemplars, the retrieved documents tended to reflect dominant academic norms. Students from multilingual or non-traditional academic backgrounds may therefore receive feedback that implicitly pushes them toward standardised rhetorical patterns, potentially limiting inclusivity and reinforcing existing academic hierarchies.

The system's inability to incorporate the affective dimensions of feedback further limits its pedagogical value. Instructors often adjust tone, encouragement, and task framing based on their understanding of a student's emotional state or learning trajectory an aspect the RAG system cannot replicate. While the model produced supportive language, it lacked the contextual sensitivity required to motivate struggling learners or respond empathetically to moments of confusion or discouragement.

Additionally, the current evaluation focused on final submissions rather than iterative draft development. Although the RAG system excelled in producing consistent, rubric-aligned feedback, it was not designed to trace learning progress across multiple revisions. As such, its potential for supporting developmental assessment which relies on tracking conceptual growth over time remains limited.

Concerns also arise regarding institutional implementation. While the tool significantly reduced marking time, the findings suggest that without appropriate training and pedagogical planning, educators may struggle to integrate RAG feedback meaningfully into classroom practice. There is a risk that institutions may misinterpret efficiency gains as justification for reducing human involvement, undermining the collaborative teacher–AI model that the system is designed to support.

Finally, student trust emerged as a crucial factor. Some students questioned how feedback was generated, what sources it relied upon, and whether human oversight was involved. These observations underscore the need for transparency and clear communication to ensure that learners understand the role of AI within assessment processes.

## 5.1 Pedagogical Potential and Advantages

Despite these limitations, the results demonstrate that the RAG-based assessment system offers significant pedagogical benefits. One of the system's strongest contributions is its capacity to deliver timely, consistent feedback at scale. In a course with over 700 submissions, the system produced detailed, rubric-aligned feedback within minutes an outcome that would be infeasible through human marking alone. This immediacy supports rapid cycles of reflection and revision, particularly when embedded within iterative writing tasks or formative assessment structures.

The system also enhanced feedback quality by retrieving high-quality exemplars and instructor-approved comments. The results indicate that students benefited from exposure to clear models of successful writing, enabling them to better understand disciplinary expectations and improve their academic literacy. When paired with intentional teaching strategies such as in-class discussions of exemplars this feature has strong potential to deepen students' understanding of assessment criteria.

From an instructional perspective, the RAG system reduced the cognitive load associated with repetitive marking tasks, enabling educators to focus on higher-order pedagogical activities. The structured outputs such as recurring weaknesses identified across the cohort provided instructors with valuable insights for curriculum development and targeted teaching interventions. Importantly, as instructors reviewed and approved system-generated feedback, these refinements were reintegrated into the knowledge base, enabling continuous improvement and alignment with evolving teaching goals.

Overall, the results suggest that the RAG-based system should be understood not as a replacement for educators, but as a pedagogical partner that enhances consistency, scalability, and feedback quality while supporting human judgement and instructional expertise.

## 6. Conclusion and Future Work

This study presents a GenAI-powered, Retrieval-Augmented Generation (RAG) system designed to enhance formative assessment in large-scale higher education contexts. By combining a large language model with a structured retrieval component accessing rubrics, exemplar essays, and instructor feedback, the system generates reliable, contextually grounded scores and qualitative feedback. Evaluation of 701 student essays demonstrated high alignment with human markers, achieving 94–99% consensus on scoring and feedback quality. These results indicate that the system can provide scalable, timely, and pedagogically meaningful assessment, supporting students' self-regulated learning and offering concrete examples of disciplinary expectations to improve academic performance.

The RAG-based approach also reduces instructor workload while maintaining consistency and transparency in assessment. By integrating prior instructor-approved feedback into its knowledge base, the system continuously improves, fostering adaptive and contextually informed guidance that complements human judgment. Students benefit from immediate, rubric-aligned feedback, enabling iterative learning cycles and deeper engagement with assessment criteria.

However, limitations remain. The system cannot fully replicate the holistic, tacit judgment of experienced instructors, including affective and dialogic dimensions of feedback. There is also a potential risk of students producing formulaic responses if they over-rely on AI-generated guidance. Additionally, the current implementation primarily evaluates final essay submissions, limiting insight into iterative learning processes and long-term skill development.

Future work should focus on expanding the system's capability to support draft-based, iterative assessment, incorporating mechanisms for guided reflection, dialogue, and affective engagement. Investigating applications across diverse disciplines, including humanities and social sciences, will help evaluate the generalizability of RAG systems. Further research could explore how adaptive AI feedback influences equity, inclusivity, and learning outcomes across heterogeneous student populations. By advancing these directions, RAG-based systems have the potential to complement human expertise, providing scalable, adaptive, and high-quality formative assessment that enriches both teaching and learning in higher education.

**Authorship contribution statement**

**Declaration of generative AI and AI-assisted technologies in the writing process**

During the preparation of this work the author used Microsoft Copilot and Open AI's ChatGPT in order to improve the readability and language of some sentences. After using these tools, the author reviewed and edited the content as needed and take full responsibility for the content of the published article.

**Funding**

This research did not receive any specific grant from funding agencies in the public, commercial, or not-for-profit sectors.

**Declaration of interests**

The author declares that there are no conflicts of interest or competing financial interests related to this study.

**Data availability**

The data and code used in this study are available upon reasonable request.